\documentclass[sigconf]{acmart}
\usepackage[inline]{enumitem}

\usepackage{microtype}
\usepackage{graphicx}
\usepackage{booktabs} 
\usepackage{multirow}

\usepackage{amsmath}
\usepackage{amsthm}
\usepackage{amsfonts}
\usepackage{mathtools}
\usepackage{bm}

\newcommand{\N}{\mathcal{N}}

\newcommand{\R}{\mathbb{R}}

\newcommand{\x}{\bm{x}}

\def\ie{\emph{i.e.}}
\def\eg{\emph{e.g.}}

\usepackage{cleveref}

\newcommand{\w}{\bm w}
\renewcommand{\x}{\bm x}

\renewcommand{\N}{\mathcal{N}}
\newcommand{\us}{DMSG}
\newenvironment{inline}{\begin{enumerate*}[label=\emph{(\roman*)}]}{\end{enumerate*}}

\AtBeginDocument{%
  }

\copyrightyear{2025}
\acmYear{2025}
\setcopyright{acmlicensed}\acmConference[RecSys '25]{Proceedings of the Nineteenth ACM Conference on Recommender Systems}{September 22--26, 2025}{Prague, Czech Republic}
\acmBooktitle{Proceedings of the Nineteenth ACM Conference on Recommender Systems (RecSys '25), September 22--26, 2025, Prague, Czech Republic}
\acmDOI{10.1145/3705328.3748072}
\acmISBN{979-8-4007-1364-4/2025/09}

\begin{document}

\title{Prompt-to-Slate: Diffusion Models for Prompt-Conditioned Slate Generation}
\author{Federico Tomasi}
\email{federicot@spotify.com}
\affiliation{%
  \institution{Spotify}
  \city{London}
  \country{United Kingdom}
}

\author{Francesco Fabbri}
\email{francescof@spotify.com}
\affiliation{%
  \institution{Spotify}
  \city{Barcelona}
  \country{Spain}
}

\author{Justin Carter}
\email{jcarter@spotify.com}
\affiliation{%
  \institution{Spotify}
  \city{NY}
  \country{US}
}

\author{Elias Kalomiris}
\email{eliask@spotify.com}
\affiliation{%
  \institution{Spotify}
  \city{NY}
  \country{US}
}

\author{Mounia Lalmas}
\email{mounial@spotify.com}
\affiliation{%
  \institution{Spotify}
  \city{London}
  \country{United Kingdom}
}

\author{Zhenwen Dai}
\email{zhenwend@spotify.com}
\affiliation{%
  \institution{Spotify}
  \city{London}
  \country{United Kingdom}
}

\begin{abstract}
\emph{Slate generation} is a common task in streaming and e-commerce platforms, where multiple items are presented together as a list or ``slate''. Traditional systems focus mostly on item-level ranking and often fail to capture the coherence of the slate as a whole. A key challenge lies in the combinatorial nature of selecting multiple items jointly. To manage this, conventional approaches often assume users interact with only one item at a time, assumption that breaks down when items are meant to be consumed together.

In this paper, we introduce \us{}, a {generative framework based on diffusion models} for prompt-conditioned slate generation. \us{} learns high-dimensional structural patterns and generates coherent, diverse slates directly from natural language prompts. Unlike retrieval-based or autoregressive models, \us{} models the joint distribution over slates, enabling greater flexibility and diversity.

We evaluate \us{} in two key domains: music playlist generation and e-commerce bundle creation. In both cases, \us{} produces high-quality slates from textual prompts without explicit personalization signals. Offline and online results show that \us{} outperforms strong baselines in both relevance and diversity, offering a scalable, low-latency solution for prompt-driven recommendation. A live A/B test on a production playlist system further demonstrates increased user engagement and content diversity.
\end{abstract}


\maketitle
\section{Introduction}\label{sec:intro}
Slate generation is an increasingly important research area, designed for scenarios where multiple items are recommended together, such as in streaming platforms like YouTube and e-commerce sites like Amazon. Unlike traditional ranking-based approaches that evaluate each candidate individually with respect to a query, slate generation focuses on selecting a group of items that collectively maximize user satisfaction~\cite{Sunehag2015-qb,Ie2019-po}. This involves optimizing the entire slate of items shown to a user accounting for factors like co-consumption patterns and diversity across items. 



Increasingly, real-world systems are being tasked with generating slates in response to prompts, themes, or editorial intents, and often without access to user interaction data. This shift is driven in part by the rise of language-based interfaces, which require models that can flexibly respond to natural language input. However, most existing methods, especially retrieval- and ranking-based  approaches, lack the generative flexibility needed in these scenarios. As a result, there is growing interest in models capable of generating entire slates from scratch, while preserving the structural coherence that makes slates meaningful.

A major challenge in slate generation lies in its combinatorial nature: jointly optimizing all items in a slate causes the number of possible item combinations to grow exponentially. Traditional methods often address this by assuming that users interact with only a single item from the slate, treating subsequent interactions as separate but conditionally dependent events~\cite{Ie2019-po}. While this assumption simplifies modeling and reduces computational complexity, it overlooks scenarios where users engage with multiple items simultaneously. This limitation is particularly evident in applications like music playlist creation~\cite{tomasi2023automatic} or bundle recommendations~\cite{sun2022revisiting,sun2024revisiting}, where items are naturally consumed together.

We propose a \emph{prompt-based slate generation model}, where the objective is to generate a slate of items conditioned solely on a natural language prompt. This framework addresses settings where historical user data is unavailable, insufficient, or used indirectly, for example by incorporating user preference in a textual format.

We formulate slate generation as a generative modeling problem and employ \emph{diffusion models (DMs)} to sample coherent and diverse slates from a learned distribution, conditioned on a prompt.
DMs are well-suited to this problem: they excel at modeling complex structures in high-dimensional spaces and generating diverse, high-quality outputs~\cite{ho2020denoising,sohl2015deep}.
Our novel application of DMs to prompt-based slate generation eliminates the need for explicit combinatorial optimization. Instead, the model learns the implicit structure of desirable slates from the data, enabling the direct generation of item sets directly from prompts.

A key strength of this approach lies in its stochasticity-driven diversity: the model can generate multiple high-quality, yet distinct slates for the same prompt. This capability can enhance content freshness and promote the discovery of less popular items--an especially valuable trait in large catalogs where many relevant combinations may exist. Our evaluations show that this generation process improves both relevance and diversity, ultimately enhancing the user discovery experience~\cite{diaz2020evaluating}.


Our framework, \emph{Diffusion Model for Slate Generation} (\textbf{\us}), generates a distribution of slates conditioned on a natural language prompt. 
To support dynamic, large-scale catalogs with millions of items, we represent items as embeddings in a continuous space and train the diffusion model on concatenated item embeddings. At inference time, the generated latent representations are decoded into catalog items using a decoding module. This design ensures that the core model remains stable as the catalog evolves—only the item mapping function needs updating, making it adaptable to real-world deployments. 
Moreover, this approach enables efficient training through the use of closed-form lower bounds~\cite{rombach2021high,li2022diffusion}. To ensure the system is suitable for low-latency, real-time recommendation, we incorporate DDIM~\cite{song2020denoising}, a noise scheduler that significantly accelerates the generation process while maintaining output quality.

We evaluate \us{} across two key tasks: music playlist generation and e-commerce bundle creation. Our model consistently outperforms state-of-the-art retrieval-based baselines, achieving improvements of up to $+12.9\%$ in average precision and $+17\%$ in NDCG over the next-best model. \us{} also produces high-quality recommendations, with an average BertScore of $0.8$, while delivering strong content freshness. 
To validate the approach in a real-world environment, we ran a two-week live A/B test on a production music playlist generation system. Results demonstrated tangible benefits: a $-13.4\%$ reduction in duplicated recommended content and a $+6.8\%$ increase in active user interactions, highlighting both the quality and impact on users of prompt-based slate generation.

The remaining of this paper is structured as follows. 
\Cref{sec:related-work} reviews related work and the state of the art is slate generation and diffusion models.
\Cref{sec:problem-formulation} formalizes the slate generation problem. 
\Cref{sec:background,sec:methodology} introduce our methodology based on continuous diffusion models. 
\Cref{sec:experiments} presents our offline experiments, while \Cref{sec:online-experiments} reports on our online evaluation. Finally, 
\Cref{sec:conclusion} discusses key contributions and outlines directions for future work.

\section{Related Work}\label{sec:related-work}

We review relevant literature, highlighting key methodologies and findings, as well as the gaps our work aims to address.

\paragraph{Slate Generation.}
Recommendation systems are typically trained to present users with the most relevant items, often as a ranked list. In such lists, top-ranked items receive disproportionate attention due to positional bias~\cite{craswell2008experimental}. 
However, in applications like music playlist creation~\cite{singh2021neural,chang2021music} or e-commerce bundles~\cite{sun2024revisiting}, users are expected to engage with most or all recommended content. In these cases, the combination of items, known as a slate, becomes essential to optimize. 
This is known as the \textit{slate generation} problem~\cite{Sunehag2015-qb,Ie2019-po,tomasi2023automatic}.

Several approaches have been proposed to address slate generation. Reinforcement learning (RL) is a common strategy~\cite{Ie2019-po}, though it often depends on simplifying assumptions, such as modeling users as selecting only the ``best'' item from a slate, to reduce the combinatorial complexity. Some methods employ user preference models to guide slate construction and train RL agents~\cite{tomasi2023automatic}. 
However, RL-based methods typically require extensive interaction data and strong modeling assumptions, limiting their practicality.
Recent work explores transformer-based models that bypass RL altogether~\cite{chen2021decision}. While prior work largely focuses on personalization using historical interactions, our approach diverges by generating slates from scratch using only natural language prompts, enabling applicability even in settings where personalization is infeasible.

\paragraph{Generative Modeling.}
%
Generative models, particularly large language models (LLMs), have shown remarkable generalization and output quality across language-centric tasks. Transformer-based, decoder-only architectures~\cite{brown2020language} have become standard and are increasingly adapted for recommendation~\cite{lin2022survey}. These models offer a compelling alternative to RL, as they do not require predefined reward functions or suffer from slow convergence. They also incorporate textual context, such as user queries or profile descriptions.

While LLMs are adaptable for recommendation, they often fall short in large-scale retrieval tasks~\cite{cao2024aligning} due to mismatches between their general-purpose knowledge and the domain-specific nature of catalogs. In particular, LLMs require prior exposure to catalog items or external mappings between textual and catalog representations~\cite{cao2024aligning}. This makes them less practical in dynamic real-world settings where catalogs frequently change and models must scale to unseen items~\cite{zhai2024actions}.
In contrast, our approach conditions generation on text prompts but employs diffusion models rather than LLM-style decoding. Diffusion models are better suited for learning the structure of high-cardinality, dynamic item spaces, making them more robust and scalable for real-world deployment.

%
%

\paragraph{Diffusion Models for Recommendations.}
%
Diffusion models (DMs)~\cite{rombach2021high} are a powerful class of generative models capable of producing structured, high-dimensional outputs, such as images or audio, from text prompts. Their ability to generalize to out-of-distribution prompts, coupled with their inherent stochasticity, makes them an attractive option for recommendation tasks.
Recent studies show that DMs can outperform collaborative filtering and traditional retrieval methods by generating diverse, coherent user-item interaction pairs~\cite{wang2023diffusion}. Some efforts have extended DMs to sequential recommendation~\cite{lin2023discrete,yang2024generate}, where models generate the next item in a user sequence based on historical data. However, these methods still rely heavily on personalized signals, limiting their applicability in data-sparse settings.

%


Our work is the first to apply diffusion models to prompt-based slate generation, where full item slates are generated from scratch using only natural language input. This enables our model to operate effectively in cold-start and editorial settings, where personalization is not feasible or desired. By modeling the joint distribution over entire slates, rather than individual items, our approach captures higher-order dependencies and enables flexible, diverse slate generation from a single prompt. 

This expands the role of diffusion models beyond personalized sequence generation to broader applications like content curation, editorial tooling, and theme-based recommendation—especially in dynamic environments with limited behavioral data~\cite{zhai2024actions}.


%

\paragraph{Diversity in Recommender Systems.}
Diversity has become a key concept in recommendation systems, often considered as important as relevance~\cite{vargas2011rank}. Prior work shows that exposure to diverse content correlates with long-term user retention~\cite{anderson2020algorithmic}, while low diversity can lead to filter bubbles and echo chambers~\cite{srba2023auditing, robertson2009probabilistic}.
Recent studies argue for treating diversity not merely as a list-level metric but as a mechanism for user exploration and knowledge discovery. For example, \citet{coppolillo2024relevance} model recommendation as a process of knowledge acquisition, where diverse recommendations maximize informational gain.
Parallel efforts have explored using LLMs for diversity-aware re-ranking. \citet{carraro2024enhancing} show that LLM-based re-rankers can improve diversity over standard ranking pipelines, though they often underperform compared to specialized models. These findings suggest that generation-based approaches may play an increasingly important role in diversity-aware recommendation.

Our method contributes to this line of work by using diffusion models to stochastically generate slates rather than re-ranking fixed outputs. This generative process naturally yields slates that are both relevant and diverse, as multiple valid item combinations can be sampled from the model's learned distribution. As shown in our experiments, this leads to high-quality recommendations that significantly improve content freshness and variety, without requiring explicit diversity constraints~\cite{kaminskas2016diversity}.

\section{Problem Formulation}\label{sec:problem-formulation}
We address the problem of slate generation, defined as selecting a list (or slate) of items to present to a user with the goal of maximizing user satisfaction. In our offline evaluation, we quantify user satisfaction through two key metrics: relevance and diversity. Slate generation plays a central role in many practical recommender systems, including bundle recommendation~\cite{avny2022bruce,sun2024revisiting}, video recommendation~\cite{Ie2019-po}, and music playlist generation~\cite{singh2021neural,tomasi2023automatic}. The objective in these settings is to construct the most effective combination of items under the assumption that users will engage with a substantial portion (or the entirety) of the recommended slate. 

We focus on two example domains: music playlist generation and e-commerce bundle recommendations. 
We refer to ``catalog items'' as the basic units of recommendation, which could be music tracks in a playlist or products on an e-commerce platform. 
Catalogs often span thousands to millions of items, with new ones being continuously added or removed, especially in dynamic environments like streaming services and online retail platforms.

We assume access to data in the form of item sequences, each paired with a textual description (\textit{labels}). 
Each item is associated with metadata, such as title, artist, and release date for music, or category and product information in e-commerce. 
The task is to generate a slate of items that aligns with a text query, optimizing user satisfaction without relying on historical user interaction data.

We formulate the problem using continuous diffusion processes (DMs)~\cite{rombach2021high}, inspired by their demonstrated ability to generate high-quality, structured content such as images and audio. This motivates our exploration of their potential in the context of slate generation.

\section{Preliminaries}\label{sec:background}

Diffusion models~\cite{ho2020denoising} are probabilistic generative models based on Markov chains, trained to learn a data distribution $q(\bm x_0)$ where $\x_0 \in \R^d$. These models generate samples through a finite sequence of steps, ultimately producing an output distribution $p_\theta(\bm x_0) = \int p_\theta(\bm x_{0:T}) d\bm x_{1:T}$, where $\bm x_1, \dots, \bm x_T$ are latent variables with the same dimensionality as the data $\bm x_0 \sim q(\bm x_0)$. 
DMs consist of two key components: the forward and the reverse process. 
The forward process, also known as the diffusion process, is a fixed, non-learned Markov chain $q(\x_{1:T} | \x_0)$ that progressively adds Gaussian noise to the input data over time. This process is governed by a predefined variance schedule $\beta_1, \dots, \beta_T$ and is defined as:
\begin{align*}
q\left(\x_{1: T} \mid \x_0\right)&:=\prod_{t=1}^T q\left(\x_t \mid \x_{t-1}\right), \\
q\left(\x_t \mid \x_{t-1}\right)&:=\mathcal{N}\left(\x_t ; \sqrt{1-\beta_t} \x_{t-1}, \beta_t \mathbf{I}\right)    
\end{align*}
The forward process incrementally adds a fixed amount of noise at each step, progressively diminishing the signal from the original data. Importantly, sampling $\x_t$ at step $t$ can be done in closed form: 
$$q\left(\x_t \mid \x_0\right)=\mathcal{N}\left(\x_t ; \sqrt{\bar{\alpha}_t} \x_0,\left(1-\bar{\alpha}_t\right) \mathbf{I}\right)$$
where $\alpha_t := 1- \beta_t$ and $\bar{\alpha}_t := \prod_{s=1}^t \alpha_s$. Using the reparametrization trick, we can sample $x_t$ in closed form as:
$\x_t = \sqrt{\bar{\alpha}_t} \x_0 + \sqrt{1-\bar{\alpha}_t}\epsilon_t$, where $\epsilon_t \sim \N(\bm 0, \bm I)$.

The reverse process models the joint distribution $p_\theta(\bm x_{0:T})$ as a Markov chain with Gaussian transitions, starting from $p(\x_T) = \N(\x_T; \bm 0, \bm I)$. In practical terms, this means that DMs gradually denoise a normally distributed sample $p(\x_T)$ over $T$ steps to recover a sample from the data distribution. The full reverse process is defined as:
\begin{align*}
p_\theta\left(\x_{0: T}\right)&:=p\left(\x_T\right) \prod_{t=1}^T p_\theta\left(\x_{t-1} \mid \x_t\right),\\
p_\theta\left(\x_{t-1} \mid \x_t\right)&:=\mathcal{N}\left(\x_{t-1} ; \boldsymbol{\mu}_\theta\left(\x_t, t\right), \boldsymbol{\Sigma}_\theta\left(\x_t, t\right)\right),
\end{align*}
where $\boldsymbol{\mu}_\theta$ and $\boldsymbol{\Sigma}_\theta$ are functions parameterized by neural networks. Following from \cite{sohl2015deep}, we train only $\boldsymbol{\mu}_\theta$ and set $\boldsymbol{\Sigma}_\theta = \sigma_t^2\bm I$ using a small fixed variance.

Training is performed by optimizing the variational bound on the negative log likelihood, 
which can be further simplified by ignoring the variance terms and edge effects, as shown in~\cite{ho2020denoising}:
\begin{align}\label{eq:logl-simple}
L_{\text {simple }}(\theta):=\mathbb{E}_{t, \mathbf{x}_0, \boldsymbol{\epsilon}}\left[\left\|\boldsymbol{\epsilon}-\boldsymbol{\epsilon}_\theta\left(\sqrt{\bar{\alpha}_t} \mathbf{x}_0+\sqrt{1-\bar{\alpha}_t} \boldsymbol{\epsilon}, t\right)\right\|^2\right]
\end{align}
where $t\sim \mathcal{U}(1,T)$ is uniformly sampled between $1$ and $T$. This approach stabilizes training, making it more efficient and leading to higher-quality generation. The network $\bm\epsilon_\theta$ is trained to approximate the true reverse distribution $p_\theta$. Different parametrizations of $p_\theta$ are possible, each corresponding to different loss functions.
In our model, we adopt the \textit{v-prediction}~\cite{salimans2022progressive}. Specifically, we predict the velocity value $\bm v_t = \alpha_t\epsilon_t - \sigma_t\x_t$, 
with a slight modification of the original loss function:
\begin{align}\label{eq:logl-simple2}
L_{\text{\us{}}}(\theta):=\mathbb{E}_{t, \mathbf{x}_0, \boldsymbol{v}}\left[\left\|\boldsymbol{v}-\boldsymbol{v}_\theta\left(\sqrt{\bar{\alpha}_t} \mathbf{x}_0+\sqrt{1-\bar{\alpha}_t} \boldsymbol{\epsilon}, t, c\right)\right\|^2\right],
\end{align}
where $c$ represents the conditional information.
This formulation correspond to ``SNR+1'' weighting~\cite{salimans2022progressive}. Empirically, we found that this loss function leads to better convergence and overall training performance compared to directly predicting $\bm\epsilon$ or $\x$.

Once the model $p_\theta$ is trained, the diffusion process can be used to generate new samples starting from $\bm x_T \sim \N(\bm 0, \bm I)$, and then iteratively sampling $\bm x_{t-1} \sim p_\theta(\bm x_{t-1} | \x_t)$ for $t\in [0, T]$.

\section{Methodology}\label{sec:methodology}
\begin{figure}
    \centering
    \includegraphics[width=.85\linewidth]{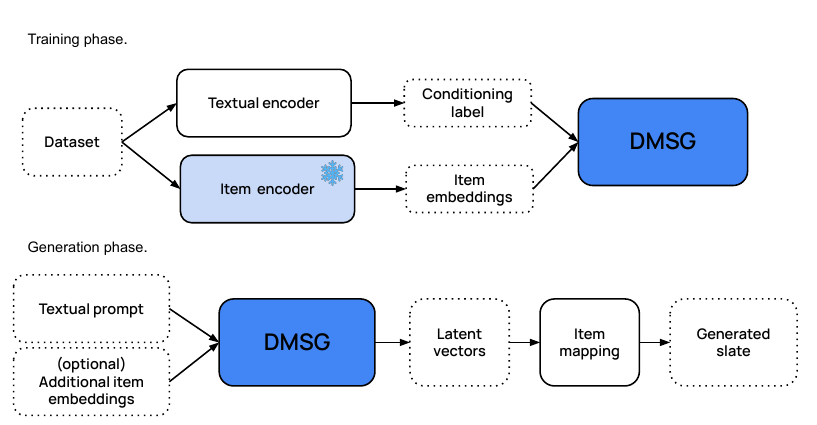}
    \caption{Overview of DMSG during training (top) and inference (bottom) phases.}
    \label{fig:dmsr}
\end{figure}
Slate generation refers to the task of sampling an ordered list of items from a potentially large and dynamic catalog.
Each slate consists of items grouped by shared characteristics and may vary in length. For simplicity, we standardize the input by padding sequences to a fixed length and discarding items from slates that exceed this maximum size. 
Each slate is also paired with a textual description that serves as the conditioning context. 

Our model, which we call the Diffusion Model for Slate Generation (\us{}), is trained to reconstruct slates based on this conditioning input. 
An overview of \us{} is shown in \Cref{fig:dmsr}.
%
\us{} consists of three main components:
\begin{inline}
\item an \textit{encoding module}, which maps discrete catalog items into a continuous latent space; 
\item a \textit{conditioning module}, which encodes the textual context into a fixed-length vector; and 
\item a \textit{diffusion process module}, a probabilistic generative model that iteratively transforms a latent vector into Gaussian noise and back, enabling the generation of coherent item sequences.
\end{inline}

\paragraph{Encoding Module.}\label{sec:method-encoding}
We assume access to a discrete sequence $\w$ of length $n$, where each $w_i \in \mathcal{V}$ is an item selected from a catalog $\mathcal{V}$ (also referred to as \textit{vocabulary}), similarly to text generation setting. Importantly, we do not assume a fixed vocabulary; instead, the catalog can dynamically change during generation.

To use diffusion models, we convert the discrete sequence $\w$ into a continuous representation $\x$ in a low-dimensional latent space~\cite{li2022diffusion}.
This is achieved through an embedding function $\phi$, which maps each item $w_i$ to a continuous vector in $\x_i\in \R^d$. 
The full sequence embedding is thus $\phi(\w) = \{ \phi(w_1), \dots, \phi(w_n) \}$, where $\phi(\w) \in \R^{n\times {d_\phi}}$. This process is flexible and can be implemented in various ways to support different types of input data~\cite{mikolov2013efficient,liang2020anchor}. 

In alignment with the continuous diffusion model formulation described in \Cref{sec:background}, we define the latent encoding of each item $w_i$ as $\x_i = \phi(w_i)$. This enables the direct use of diffusion models in the continuous domain, where $\x_i$ serves as the representation of catalog item $w_i$. 

We explored two types of encoding modules: 1) a trainable encoder, jointly optimized with the diffusion model, and 2) a fixed encoder, trained separately and held constant during the DM training.
Training the encoder jointly with the diffusion model introduces additional parameters and requires modifying the optimization objective (see \Cref{eq:logl-simple}). 
However, preliminary experiments~\cite{rombach2021high,li2022diffusion,strudel2022self} suggest that using a fixed encoder offers practical benefits, including improved training stability and comparable performance. Therefore, in this work, we adopt a fixed encoding module, leaving a comprehensive comparison with trainable alternatives for future work.

\paragraph{Conditioning Module.}
To guide the diffusion process toward generating contextually appropriate slates, we model conditional distributions of the form $p(\x|c)$, where $c$ represents the \textit{context} variable~\cite{rombach2021high}.
The context can take various forms, such as a textual description (also referred to as \textit{prompt}) or additional related catalog items. By conditioning on such information, the model can generate outputs that align with specific semantic intents~\cite{reed2016generative}

\begin{figure}[t]
    \centering
    \includegraphics[width=\linewidth]{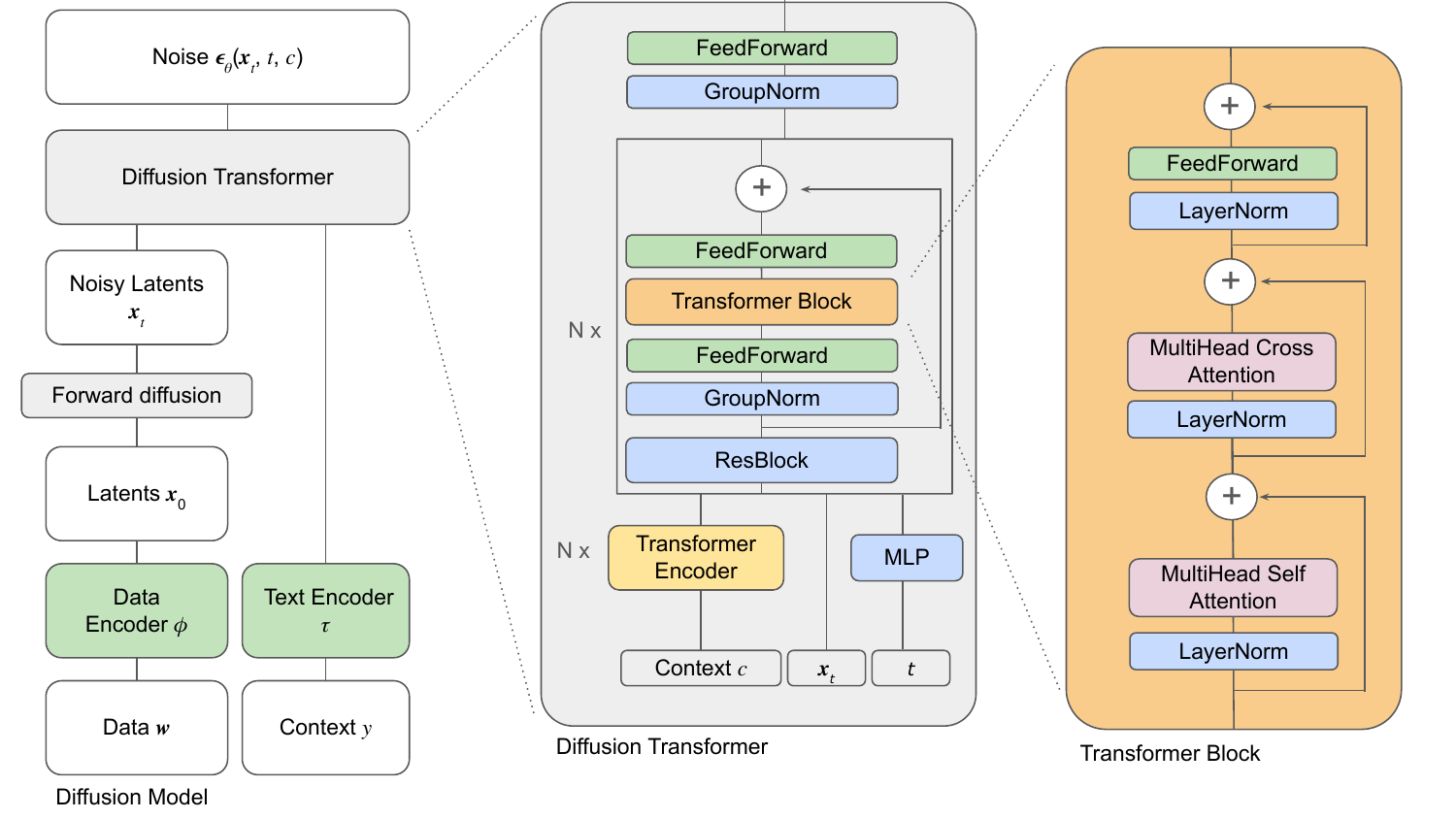}
    \caption{Diffusion model and transformer details. The core module is a diffusion transformer to predict the noise artificially added to latent variables. At generation time, this is used to denoise a latent vector.}
    \label{fig:transformer}
\end{figure}

We preprocess the conditional text using a domain-agnostic encoder $\tau$, which maps an input variable $y$ (\eg, a textual prompt) into a vector representation $\tau(y) \in \R^{M \times d_\tau}$. For simplicity, we refer to this encoded vector as context $c$.
To implement $\tau$, we use a stack of transformer modules, well suited for capturing the structure and semantics of textual inputs~\cite{vaswani2017attention}. 
The resulting context vector $c$ is used to condition the diffusion process via a cross-attention mechanism within the transformer-based architecture. \Cref{fig:transformer} shows a detailed illustration of DMSG and its components.

\paragraph{Diffusion Process.}
At the core of our model is a conditional diffusion process--a probabilistic generative model that iteratively transforms a random sample drawn from a Gaussian distribution into a structured sample from a target distribution, represented in a compact latent space~\cite{sohl2015deep}. As described in \Cref{sec:background}, the diffusion process consists of two components: a forward, fixed components that gradually adds noise to the latent variables, and a learnable reverse process that removes this noise to recover the original signal. 
To implement the reverse process, we use a transformer-based architecture designed to capture sequential dependencies within the slate and enable scalable training. This design follows recent successful applications of transformers in image generation tasks~\cite{peebles2023scalable}. We refer to this core component as the \textit{diffusion transformer}. 

The diffusion transformer incorporates a cross-attention mechanism within its core blocks~\cite{vaswani2017attention}, similar to latent DMs that condition generation on external information such as class labels and textual prompts~\cite{rombach2021high}.
In our setup, the transformer processes the context vector $c$, the noisy latent input $\x_t$ and the diffusion step $t$, and is trained to predict the noise component originally added to the clean sample $\x_0$.
This module, denoted $\bm v_\theta(\x_t, t, c)$, predicts the \textit{velocity} (described in \Cref{sec:background}),  and is ultimately used to reconstruct the original latent sample~$\x_0$.


At generation time, the model takes the conditioning information, denoted as $y$, typically in the form of natural language. The input is processed by the text encoder $\tau$, producing a latent representation $c = \tau(y)$.  To initiate the generation process,  we sample an initial latent vector $\x_T\sim\N(\bm 0, \bm I)$ from a standard Gaussian distribution and iteratively apply the reverse diffusion process (or a subset of its steps) to generate a new sample. Each step is conditioned on $c$, ensuring that the final output reflects the original prompt $y$.
Formally, the reverse process at each timestep computes:
$$\x_{t-1} = \frac{1}{{1 - \bar{\alpha}_t}} \left(\sqrt{\bar{\alpha}_{t-1}} \beta_t\hat{\x}_0  + \sqrt{\alpha_t} (1 - \bar{\alpha}_{t-1}) \x_t\right) $$
where the predicted clean sample $\hat{\x}_0$ is given by:
$$\hat{\x}_0 = \sqrt{\bar{\alpha}_t} \x_t - \sqrt{1-\bar{\alpha}_t} \bm v_\theta(\bm x_t,t,c).$$
After the final step, the continuous representation $\bm x_0$ is converted into a discrete slate  $\w$ using a fixed (non-learned) \emph{rounding} step, parametrized by $p_w(\w | \x_0)$~\cite{li2022diffusion}.

This decoding step enables flexibility in handling dynamic catalogs at inference time, as new items can be added or removed without retraining the model. By simply updating the mapping function  $p_w(\w | \x_0)$, the system can adjust to changes in the item space. For example, to avoid recommending duplicate items within a single session, small variations can be introduced to $\bm x_0$. 
If a nearest neighbor mapping is used, it may yield the same items repeatedly; in such cases, duplicate recommendations can be masked, and the next nearest item in the embedding space can be selected instead.

\paragraph{Noise Scheduler}
A well-known limitation of DMs is the need to sequentially iterate through all  $T$ diffusion steps. While using a large $T$ can improve training stability and sample quality, it also requires the reverse diffusion process to traverse each step $t \in \{1, \dots, T\}$, often numbering in the thousands. This makes generation computationally expensive, posing challenges for low-latency, user-facing applications.

To address this,  we leverage fast sampling techniques, specifically 
denoising diffusion implicit models (DDIMs)~\cite{song2020denoising}. DDIMs generalize the Markovian diffusion process of DDPM to a non-Markovian formulation that retains the same training objective. This enables us to skip intermediate steps during inference, bypassing the full step-by-step trajectory used during training.
Using DDIMs, we achieve fast generation times (on the order of milliseconds per sample) while maintaining comparable sample quality. In practice, we reduce the number of inference steps to 50, allowing us to meet the strict latency requirements of real-time recommendation systems.




\section{Offline Experiments}\label{sec:experiments}


We evaluate \us{} on two distinct slate generation tasks: music playlist generation and e-commerce bundle recommendation. Our experiments are conducted using a subset of (prompt, slate) pairs, and assume that all information about the relevance of the slate is encapsulated within the prompt itself.

\subsection{Experimental setup}\label{setup}

\paragraph{Baselines.}
We compared our diffusion-based approach with several unpersonalized baselines that also generate recommended slates from textual prompts:
\begin{enumerate*}
  \item \textbf{Popularity}: A heuristic that recommends the most popular items in the dataset, independent of the input prompt. Popoularity is computed using the frequency of items within the dataset.
  \item \textbf{Prompt2Vec}: An encoder-only model that maps a sentence embedding of the input query through two fully connected layers to produce a dense vector. Similar to \us{}, we retrieve the closest catalog items using a nearest neighbor mapping in the embedding space. The model is trained at the item level using associated metadata as input.
  \item \textbf{BM25}: A traditional retrieval method that scores items based on their textual relevance to the prompt, using the BM25 ranking function~\cite{robertson2009probabilistic}.
  \item \textbf{S2S}: A sequence-to-sequence model built on a transformer architecture~\cite{vaswani2017attention}, consisting of an encoder-decoder pair. The encoder processes the textual prompt, while the decoder generates a probability distribution over the item vocabulary, from which we select the top candidates for each prompt to generate the slate.
\end{enumerate*}

\paragraph{Datasets.}
We evaluate our approach on the following datasets:
\begin{enumerate*}
\item \textbf{Spotify Million Playlist Dataset (MPD)}: This dataset contains 1 million playlists created by Spotify users between January 2010 and October 2017, and include playlist title, description, and and ordered lists of tracks with corresponding artist and track names.\footnote{\url{aicrowd.com/challenges/spotify-million-playlist-dataset-challenge}} For training and evaluation efficiency, we randomly sample a subset of 100K playlists for our experiments.
\item \textbf{Curated playlists (Curated)}: A proprietary dataset from a popular music streaming platform, comprising 30,000 high-quality, editorially curated playlists. Each playlist includes the same metadata as MPD.  
\item \textbf{Bundle Recommendation Dataset (Bundle)}: Based on Amazon datasets, this dataset includes 11,000 unique catalog items that appear in 5,444 bundles~\cite{sun2022revisiting,sun2024revisiting}.\footnote{\url{github.com/BundleRec/bundle_recommendation}}

\end{enumerate*}
Each dataset is split into training and test sets using a 80-20 ratio. The split is slate-wise, where 80\% of the slates are used for training, and 20\% for the test set which is used exclusively for evaluation. All reported metrics are computed on the test set. 

\paragraph{Model training.} We trained \us{} for 1,000 epochs using the Adam optimizer~\cite{kingma2014adam} with a learning rate of $10^{-4}$. 
The velocity prediction model $\bm v_\theta$ consists of three transformer layers (\Cref{fig:transformer}), each preceded by a 320-dimensional residual block with \textit{swish} activation and group normalization (20 groups). Each transformer module includes a group normalization layer (20 groups), 10 heads with 32 dimensions each, a single transformer block between two dense layers, and a final \textit{swish} activation and group normalization layer at the end of the sequence.
We encode the context with 3 encoder-only transformer layers, 
each with 8 attention heads, a hidden dimension of 128,
and a 10\% dropout rate with a ReLU activation. 

To generate training labels for each slate, we used item metadata--including title, descriptions, and curated keywords--processed through a pretrained text encoder.\footnote{huggingface.co/ianstenbit/keras-sd2.1} This encoder supports input sequences of up to 77 tokens~\cite{rombach2021high}. This setup effectively captures the nuances of textual prompts in our target applications. However, other domains may require different encoders to accommodate longer inputs. Notably, our framework supports plug-and-play integration of arbitrary text encoders with minimal or no changes to the model architecture.

For catalog item embeddings, we set the encoding dimension to 80, with a sequence length of 32 for the Curated and MPD datasets, and 5 for the Bundle dataset. This is the maximum slate size we consider, padding smaller sequences and trimming longer sequences to such sequence length. For Curated and MPD datasets we use a Word2Vec-based encoder that encodes tracks based on co-occurrence in playlists \citep{mehrotra2018towards}, while for Bundle dataset we use a pre-trained LLM\footnote{https://huggingface.co/sentence-transformers/all-MiniLM-L6-v2} to encode item textual features in a zero-shot setting.
For comparison, we also trained the S2S and Prompt2Vec baselines for 1,000 epochs using the Adam optimizer with a learning rate of $10^{-4}$. Prompt2Vec uses a single hidden layer dimension of 4096, while S2S uses an output dimensionality of 2048 for both its decoder and encoder components. 

\subsection{Results}\label{Offline_Evaluation}

We present results evaluating the effectiveness of DMSG across our two slate generation tasks: music playlist generation and e-commerce bundle recommendation.

\subsubsection{Is \us{} effective at generating relevant slates?}
We start by evaluating the relevance of items generated by \us{} in response to a given prompt. We assume that all information needed to assess the relevance of a slate is contained within the prompt itself. This approach makes \us{} directly comparable to retrieval-based baselines such as BM25, which also rely solely on the input prompt.

To evaluate relevance, we compare \us{} against baseline models using reference oracle slates and standard ranking metrics. For each model, we generate candidate slates conditioned on the reference prompt and assess how closely the generated items match the reference slate. Catalog items are considered relevant if they appear in the reference slate, aligning with common recommender system evaluation practices~\cite{said2014comparative}. As a result the reported scores should be interpreted as lower bound estimates, since models may generate equally relevant items that are not present in the reference slate. 
\begin{table}[t]
    \centering
\resizebox{0.99\linewidth}{!}{%
\begin{tabular}{lllll}
\toprule
\textbf{Curated} & NDCGSim@30 & MAPSim@30 & BertScore & CategorySim@30 \\
\midrule
Popularity & 0.0800 (+- 0.11) & 0.1010 (+- 0.16) & 0.2952 (+- 0.23) & 0.0057 (+- 0.04) \\
Prompt2Vec & 0.0674 (+- 0.10) & 0.0762 (+- 0.14) & 0.2670 (+- 0.22) & 0.0048 (+- 0.02) \\
BM25 & 0.6054 (+- 0.23) & 0.7955 (+- 0.23) & 0.8928 (+- 0.14) & 0.4677 (+- 0.27) \\
S2S & 0.3761 (+- 0.28) & 0.4763 (+- 0.34) & 0.6531 (+- 0.31) & 0.2153 (+- 0.24) \\
DMSG (ours) & \textbf{0.7087 (+- 0.18)} & \textbf{0.8978 (+- 0.11)} & \textbf{0.9314 (+- 0.06)} & \textbf{0.4783 (+- 0.24)} \\
\bottomrule
\toprule
\textbf{MPD} & NDCGSim@30 & MAPSim@30 & BertScore & CategorySim@30 \\
\midrule
Popularity & 0.0064 (+- 0.02) & 0.0602 (+- 0.14) & 0.0657 (+- 0.16) & 0.0066 (+- 0.03) \\
Prompt2Vec & 0.0054 (+- 0.02) & 0.0289 (+- 0.06) & 0.0684 (+- 0.17) & 0.0094 (+- 0.02) \\
BM25 & \textbf{0.5614 (+- 0.15)} & 0.7373 (+- 0.13) & 0.8444 (+- 0.10) & 0.3674 (+- 0.20) \\
S2S & 0.3506 (+- 0.16) & 0.5247 (+- 0.23) & 0.7762 (+- 0.17) & 0.2337 (+- 0.21) \\
DMSG (ours) & 0.5210 (+- 0.14) & \textbf{0.8119 (+- 0.09)} & \textbf{0.8796 (+- 0.06)} & \textbf{0.4792 (+- 0.18)} \\
\bottomrule
\toprule
\textbf{Bundle} & NDCGSim@5 & MAPSim@5 & BertScore & CategorySim@5 \\
\midrule
Popularity & 0.0870 (+- 0.09) & 0.1410 (+- 0.13) & 0.1659 (+- 0.14) & 0.0822 (+- 0.15) \\
Prompt2Vec & {0.1862 (+- 0.10)} & 0.3292 (+- 0.18) & 0.3452 (+- 0.15) & 0.0678 (+- 0.11) \\
BM25 & 0.0955 (+- 0.04) & {0.4998 (+- 0.13)} & {0.5338 (+- 0.17)} & {0.1940 (+- 0.11)} \\
S2S & 0.1406 (+- 0.10) & 0.2112 (+- 0.15) & 0.2930 (+- 0.18) & 0.1600 (+- 0.21) \\
DMSG (ours) & \textbf{0.3157 (+- 0.13)} & \textbf{0.5690  (+- 0.11)} & \textbf{0.5784 (+- 0.05)} & \textbf{0.4069 (+- 0.21)} \\
\bottomrule
\end{tabular}
}
    \caption{Comparison of DMSG with baselines, including average score and standard deviation across all generated slate.}
    \label{tab:full-metrics}
\end{table}

To better account for semantically similar but non-identical items, we introduce the cosine similarity between embeddings as part of the ground truth. Specifically, for a generated item $item_i$ in position $i$, if it does not appear in the reference slate, it is assigned a relevance score based on cosine similarity to the corresponding reference item $ref_i$, \ie, $s_i = cos\_sim(item_i, ref_i)$. This approach reduces the penalty for recommending items not present in the ground truth, but still highly similar to those and, at the same time, mitigates mismatches via approximate nearest-neighbor lookups on the item embedding space.
We introduce two similarity-based metrics: NDCGSim and MAPSim. The first one assesses the quality of the slate as whole, while the latter focuses on the relevance of the retrieved items.
We also report BertScore~\cite{zhang2019bertscore}, which measures semantic similarity between the generated and reference slates using item vectors as token embeddings, and CategorySim, a category-level similarity score that weights metadata overlaps by their frequency of appearance.

\Cref{tab:full-metrics} presents the average and standard deviation of each metric across all test prompts, providing insight into the model stability. To account for the stochasticity nature of \us{}, we sample five slates per prompt and compute average scores, and then aggregate across prompts to report overall performance. 

\us{} outperforms the baselines across nearly all metrics. On the Curated dataset, \us{} shows notable improvements in NDCGSim and MAPSim (+17\% and +12.9\% over the second-best BM25), indicating both high-quality item selection and effective slate structuring. The smallest gain is seen in CategorySim (+2.3\%), which is expected since retrieval models like BM25 already perform well in category alignment.
We observe a similar trend on the MPD dataset. While BM25 slightly outperforms \us{} on NDCGSim, \us{} achieves higher MAPSim and BERTScore, suggesting that it generally recommends more semantically relevant items even if not ranked identically to the reference. 

Results on the Bundle dataset echoes the findings from the music datasets. Due to the shorter length of bundles, we compute metrics using only the first five items. \us{} continues to outperform all baselines, confirming its ability to handle both long and short slates effectively. This demonstrates the broad applicability of \us{} across diverse slate generation scenarios.

\subsubsection{Can \us{} provide a diverse selection of items?}
Diversity is a critical aspect of recommender and retrieval systems. In this work, we focus on two complementary dimensions of diversity: 1) recommending less popular items to avoid over-concentration on frequent content and 2) providing fresh and varied results across multiple generations for the same prompt, while maintaining relevance. The first aspect encourages exposure to a broader range of catalog items by reducing reliance on popularity signals. The second, tied to the notion of freshness, ensures that repeated prompts return different--but still relevant--slates over time. This ability to produce varied recommendations is known to help maintain user interest and supports discovery.


\begin{figure}[t]
    \centering
    \includegraphics[width=0.85\linewidth]{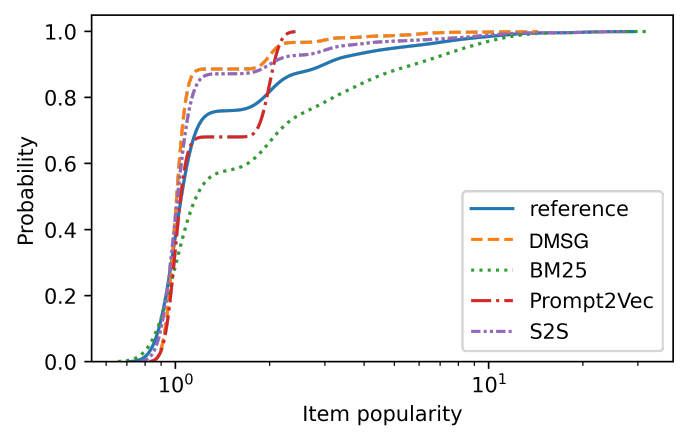}
    \caption{Cumulative distribution of exposure generated by the recommendation algorithms when compared to initial level of popularity in the Bundle dataset.} 
    \label{fig:item-popularity}
\end{figure}


\Cref{fig:item-popularity} compares the popularity distribution of items recommended by different models on the Bundle dataset, relative to the reference data. Item popularity (x-axis) is defined by the number of user interactions logged for each item, while the y-axis shows the proportion of recommended items at each popularity level. \us{}, along with S2S, tends to recommend less popular items overall, falling below the popularity level of the reference set, whereas BM25 consistently favors more popular items, often exceeding reference popularity. These results, alongside the relevance metrics in \Cref{tab:full-metrics}, show that \us{} successfully promotes lower-popularity items while maintaining high-quality slates, ultimately enabling better exploration of the catalog.

\begin{figure}[t]
    \centering
    \includegraphics[width=0.85\linewidth]{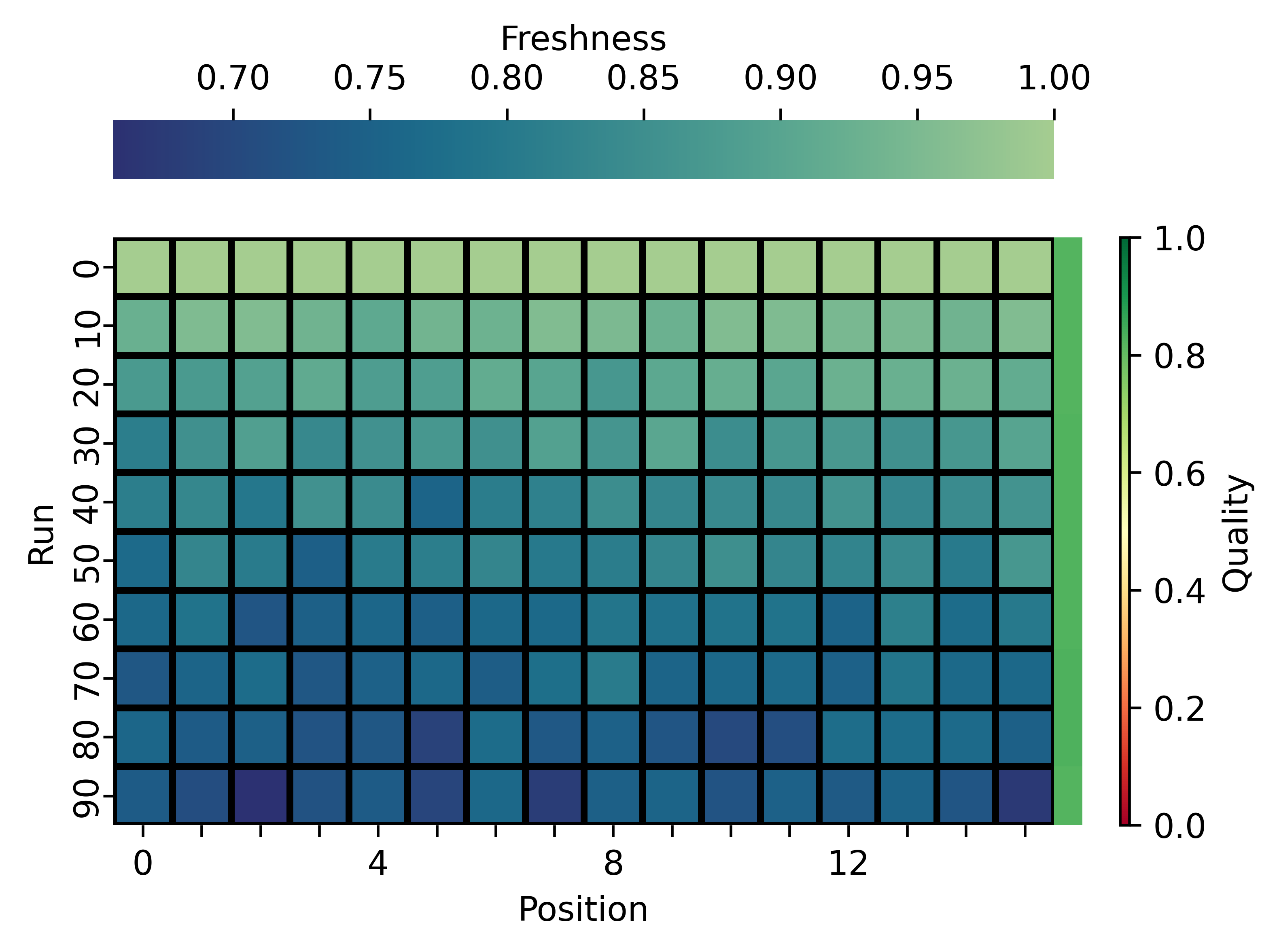}
    \caption{Repeated playlist generation with freshness and quality score. Multiple playlist from the same prompt have similar qualities and low overlap.}
    \label{fig:novelty}
\end{figure}


Another strength of DMs is their ability to naturally generate fresh content due to their stochastic sampling process in continuous latent space. We evaluate this by measuring freshness--the number of new, non-repeated items--across multiple slates generated for the same prompt. To ensure these new items are still relevant, we also measure BERTScore for each slate, assessing content quality across generations.

\Cref{fig:novelty} summarizes results on the MPD test set, showing freshness and quality across multiple runs per prompt. Each run produces a slate of items; we plot the first 16 positions for clarity. The rightmost column shows the average BERTScore per run, which remains consistently around 0.8, confirming that freshness does not come at the cost of quality.
Notably, the model surfaces new items throughout the entire slate, not just in later positions. This is evident from the spread of blue squares across all positions in the figure, indicating \us{}'s ability to inject freshness uniformly across the slate. This property is essential for delivering dynamic and engaging user experiences, as it ensures that each recommendation instance feels new, even for identical prompts.



\subsubsection{What is the effect of post-processing?}
Starting from a conditioning input, \us{} generates a slate of items in a structured order to optimize coherence and relevance based on the learned latent space. However, in real-world recommender systems, it is common to apply a post-processing step to reorder items, either to incorporate additional relevance signals or to personalize the slate for a specific user. 

To assess \us{} flexibility in such settings, we evaluate its performance both as a standalone recommender and as a candidate generator, followed by three post-processing strategies
\begin{enumerate*}[label=\emph{(\roman*)}]
    \item no post-processing, preserving the model’s original item order;
    \item sorting by item centroid, where items are ranked by proximity to the centroid of the latent slate embedding; and 
    \item random shuffling, serving as a control.
\end{enumerate*}
We focus on the MPD dataset, where the effect of post-processing is more pronounced due to the longer length of generated slates.

\begin{table}[t]
    \centering
\resizebox{0.99\linewidth}{!}{%
    \begin{tabular}{lrrrr}
\toprule
 & NDCGSim@30 & MAPSim@30 & BertScore & CategorySim@30 \\
\midrule
\us{} (random sort) & 0.5134 (+- 0.15) & 0.7330 (+- 0.10) & 0.8744 (+- 0.08) & 0.4788 (+- 0.18) \\
\us{} (no post-process) & 0.5138 (+- 0.15) & 0.7343 (+- 0.10) & 0.8744 (+- 0.08) & 0.4789 (+- 0.18) \\
\us{} (centroid sort) & \textbf{0.5210 (+- 0.15)} & \textbf{0.8119 (+- 0.09)} & \textbf{0.8796 (+- 0.08)} & \textbf{0.4792 (+- 0.18)} \\
\bottomrule
\end{tabular}
}
    \caption{Post-processing methods on MPD data.}
    \label{tab:post-processing}
\end{table}

\Cref{tab:post-processing} presents the results. Overall, the differences between variants are modest. However, sorting by item centroid slightly improves performance, suggesting that \us{} is able to position the most relevant tracks near the center of the latent distribution. By sorting these toward the top, we can ensure that the first items in the slate are most representative of the input prompt, a useful property for tasks like playlist generation, where early items often receive more attention.

\section{Online Experiments}\label{sec:online-experiments}
We evaluated our slate generation approach to assess its effectiveness in powering real-world recommender systems. 
Although trained offline on a limited subset of generic music playlists, we expected \us{} to deliver satisfactory user listening experiences. Notably, we do not update or fine-tune the model during the testing period--we leave such enhancements for future work.

\paragraph{Experiment Settings.}
To evaluate the effectiveness of our proposed approach, we conducted a live A/B test on our production platform. We compared our model, \us{}, against the platform’s default playlist generation model over a two-week period. Users were randomly assigned to one of two non-overlapping groups: the control group received playlists generated by the production model, while the treatment group received playlists generated by \us{}.

The experiment collected a large sample of interaction data, spanning 1 million users, 2.8 million unique tracks, and 4.8 million distinct sessions. Each session was divided into multiple slates, with each slate consisting of five consecutive tracks. For users in the treatment group, these tracks were generated using \us{}; for those in the control group, they were generated by the existing production model. 

The production model (control) is a personalized retrieval-based system that, given textual keywords representing the playlist's content hypothesis and user familiarity with the content hypothesis, retrieves a broader pool of relevant tracks. These are then filtered using additional user-specific features (\eg, content familiarity). Content familiarity is computed using past music tracks that the users interacted with. To ensure a fair comparison, we retained the same post-processing and sequencing pipeline across both groups, replacing only the retrieval component with \us{}, which generates the initial track set based on the same content hypothesis.

\us{} was deployed on a single NVIDIA L4 GPU system for inference. During testing, the model utilized 50 diffusion steps and consistently achieved a P99 latency of 150ms. In comparison, the baseline control system exhibited a P99 latency of 500ms. This reduction provides potential for future iterations or an overall improvement in user-perceived latency.

\paragraph{Online Evaluation.}
\begin{figure}
    \centering
    \includegraphics[width=.9\linewidth]{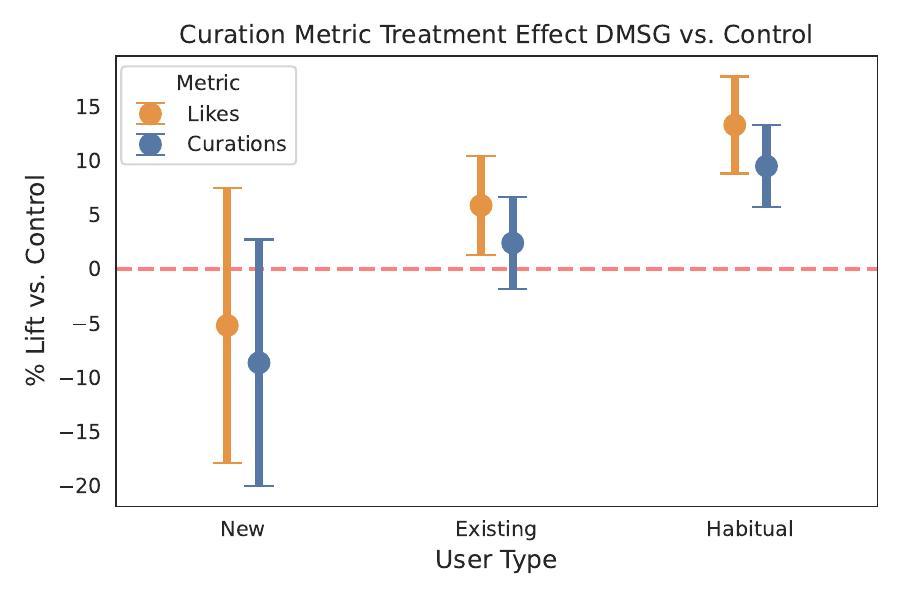}
    \caption{Differences in stream curations between treatment (\us{}) and control group divided by user type, with error bars indicating the 95\% confidence interval.}
    \label{fig:online-results}
\end{figure}

We evaluate two key performance metrics commonly used in the streaming domain: \emph{stream curations}, which capture active interaction with a track (\eg, adding it to a playlist or liking it), and \emph{listening behavior}, capturing user session listening (\eg, number of tracks skipped or minutes played).

On average, the results show a $+6.8\%$ uplift in stream curations, indicating that users engaged more actively with tracks recommended by \us{} compared to the control group. When focusing specifically on curations involving the addition of tracks to a user's ``Liked Songs'', the uplift increases to $+10.5\%$.

To better understand this effect, we break down stream curation performance by user type: 1) new users: no prior interaction on the platform before the experiment; 2) existing users: some previous but infrequent engagement; and 3) habitual users: regular, high-frequency engagement. As shown in \Cref{fig:online-results}, the uplift in stream curations is most pronounced among existing and habitual users, highlighting \us{}'s particular value for users familiar with the platform. A possible explanation is that these users, having developed some baseline familiarity or intent, may be more open to discovering new content.

Adding a track to user's Liked Songs playlist is an indication that the recommendation was both relevant and novel, as tracks already liked cannot be added again. 
While this reflects a significant improvement in active engagement, it is accompanied by a small decline in listening duration: 
$-5.6\%$ decrease in minutes played and $+3\%$ increase in skip rate (where users skipped tracks before completion). This trade-off is expected, as \us{} does not rely on explicit personalization signals and instead prioritizes content freshness, which, as discussed earlier, better supports discovery.



Importantly, active curations can serve as a proxy for how users respond to novel or previously unseen content, which is often related to content diversity. While novelty reflects a user's exposure to new items, diversity captures the broader range of content types or categories presented. The two are closely linked, as more diverse recommendations increase the likelihood of surfacing novel content. 
To directly assess content freshness, we measured the proportion of repeated tracks served to the same user during the first week post-exposure. \us{} showed a $-13.4$\% reduction in repetition compared to the control group (lower is better), indicating greater diversity in recommendations. This aligns with both our underlying assumptions and offline evaluations, highlighting how \us{}’s inherent stochastic process supports the generation of diverse, high-quality recommendations.

\section{Discussion and Future Work}\label{sec:conclusion}

We introduced \us{}, a novel method for slate generation based on diffusion models. By leveraging the generative capabilities of diffusion models, \us{} conditions on contextual information, such as textual prompts, to produce structured and coherent item slates. Its ability to generalize to unseen catalog items, combined with the stochastic nature of iterative sampling, enables \us{} to dynamically balance content freshness and relevance, making it well-suited for integration into larger recommender system pipelines.


As demonstrated in our experiments, \us{} can generate relevant item combinations from simple text prompts, opening up a range of applications. In music playlist generation, it can reconstruct reference playlists from minimal input or generate new ones based on user-provided or system-generated descriptions, such as preferred music genres. While our study focuses on text-based conditioning, our framework supports alternative conditioning signals, including catalog items or user interaction patterns.
%
In e-commerce bundle recommendations, \us{} can be paired with an external module that detects user interest in categories (\eg, ``TV accessories'') based on browsing behavior or past interactions. Users can also explicitly explore these categories, receiving fresh, dynamically updated bundles. \us{}'s stochasticity enables diverse yet relevant recommendations, supporting discovery and ultimately enhancing user satisfaction.

%

To further enhance \us{}’s capabilities, future work could explore integrating richer metadata into the generation process. Unlike pixels in images or samples in audio, catalog items often contain detailed metadata that could improve alignment between prompts and generated slates. Additionally, incorporating prompt reformulation techniques, as explored in recent studies~\cite{ma2024exploring}, may offer finer control over slate generation and improve output quality.

\bibliographystyle{ACM-Reference-Format}
\bibliography{main-bib}

\appendix

\end{document}